\documentclass{iaus}
\usepackage{graphicx}

\title[LBVs \& Mass Loss near Eddington Limit] 
{Luminous Blue Variables \& Mass Loss \\ near the Eddington Limit}

\author[S. Owocki
\& A.-J. van Marle]   
{Stan Owocki
\and Allard Jan van Marle
}

\affiliation{Bartol Research Institute, 
Department of Physics \& Astronomy \\ 
University of Delaware,  Newark, DE 19350 USA \\ 
email: {\tt owocki@bartol.udel.edu}, 
{\tt marle@udel.edu}\\
}

\pubyear{2008}
\volume{250}  
\pagerange{1--2}
\jname{Massive Stars as Cosmic Engines}
\editors{Fabio Bresolin, Paul Crowther \& Joachim Puls, eds.}
\begin{document}

\def\spose#1{\hbox to 0pt{#1\hss}}
\def\ltwig{\mathrel{\spose{\lower 3pt\hbox{$\mathchar"218$}}
     \raise 2.0pt\hbox{$\mathchar"13C$}}}
\def\gtwig{\mathrel{\spose{\lower 3pt\hbox{$\mathchar"218$}}
     \raise 2.0pt\hbox{$\mathchar"13E$}}}
\def\blankline{\par\vskip \baselineskip}
\newcommand{\beq}{\begin{equation}}
\newcommand{\eeq}{\end{equation}}
\newcommand{\beqa}{\begin{eqnarray}}
\newcommand{\eeqa}{\end{eqnarray}}

\maketitle

\begin{abstract}
During the course of their evolution, massive stars lose a substantial
fraction of their initial mass, both through steady winds and through
relatively brief eruptions during their Luminous Blue Variable (LBV)
phase.  This talk reviews the dynamical driving of this mass loss,
contrasting the line-driving of steady winds to the potential role of
continuum driving for eruptions during LBV episodes when the star
exceeds the Eddington limit.  A key theme is to emphasize the inherent
limits that self-shadowing places on line-driven mass loss rates,
whereas continuum driving can in principle drive mass up to the
``photon-tiring" limit, for which the energy to lift the wind becomes
equal to the stellar luminosity.  We review how the ``porosity" of a
highly clumped atmosphere can regulate continuum-driven mass loss, 
but also discuss recent time-dependent simulations of how base mass flux 
that exceeds the tiring limit can lead to flow stagnation and a complex,
time-dependent combination of inflow and outflow regions.  
A general result is thus that porosity-mediated continuum driving in
super-Eddington phases can explain the large, near tiring-limit mass loss
inferred for LBV giant eruptions.
\keywords{stars: early-type, stars: winds, outflows, stars: mass loss, 
stars: activity}
\end{abstract}

\firstsection 

\section{Introduction}

Two key properties in making massive stars ``cosmic engines'' are
their high luminosity, and their extensive mass loss.
Indeed the momentum of this radiative luminosity is a key factor in driving
massive-star mass loss, for example through the coupling with bound-bound 
opacity that is the basis of their more or less continuous 
line-driven stellar winds.
Among the most luminous hot stars there appears a class of ``Luminous 
Blue Variables'' (LBVs) for which the winds are particularly
strong, and exhibit irregular variability on time scales ranging 
from days to years.
Contemporary observations generally suggest modest variations in net mass
loss, occuring with nearly constant bolometric luminosity, and which
might readily be explained by, e.g., opacity instabilities within the standard
line-driving mechanism.
But historical records, together with the extensive nebulae around
many LBVs, suggest there are also much more dramatic eruptions,
marked by substantial increases in the already extreme 
radiative luminosity, and lasting for several years,
over which the net mass loss,  $0.1-10~M_{\odot}$, far exceeds what 
could be explained by line-driving.
Rather, the closeness of such stars to the Eddington limit, for which 
the radiative force from just the electron scattering continuum
would balance the force of gravity, suggests that such ``giant
eruptions'' might instead arise from {\em continuum} driving,
resulting in much higher mass loss, perhaps triggered by interior
instabilities that increase the stellar luminosity above the Eddington
limit.
 
The review here focusses on the underlying physical issues behind such
historical LBV mass loss.
One particular theme is whether such eruptions are best characterized
as {\em explosions},  or as episodes of an enhanced quasi-steady {\em wind}.
Key distinctions to be made include 
timescale (dynamic vs. diffusive),
driving mechanism (gas vs. radiation pressure),
and
degree of confinement (free expansion vs. gravitationally bound).
As detailed below, it seems the characteristics of LBV giant eruptions
require a combination of each, i.e. a quasi-steady wind driven by the
enhanced lumosity associated with a relatively sudden (perhaps
even explosive) release of energy in the interior.
But even once an enhanced, super-Eddington luminosity is established,
there remain fundamental issues of how the continuum driving can
be regulated, e.g. by the spatial ``porosity'' of the medium, and thus
lead to a mass loss that in some cases is inferred to
have an energy comparable to the radiative luminosity, representing a 
``photon-tiring'' limit.

\section{The Key to Stellar Mass Loss: Overcoming Gravity}

\subsection{Basic  Momentum and Energy Requirements for Steady Wind}

Gravity is, of course, the essential force that keeps a star together 
as a bound entity, and so any discussion of stellar mass loss must
necessarily focus on what mechanism(s) might be able to overcome this 
gravity. 
There are two aspects of this, namely 
to provide the momentum needed to reverse the inward pull of
the gravitational force, but then also to have this outward driving 
sustained by tapping into a reservoir of energy that is sufficient to 
lift the material completely out of the star's gravitational potential.

For a steady radial wind flow, momenutum balance requires that any 
acceleration in speed $v$ with radius $r$ must result from a combination
of the gradient of gas pressure with any other outward force to 
overcome the inward pull of gravity,
\beq
 v \frac{dv }{dr} =
 - \frac{GM}{r^{2}} 
 - \frac{1}{\rho} \, \frac{dP}{dr}
 + g_{out}
 \, ,
\eeq
with standard notation for, e.g., mass density $\rho$ 
and stellar mass $M$. 
The discussion below focuses on radiative forces as a key to
providing the required outward driving term $g_{out}$,
but for now, let us just consider some general properties of such
steady wind models.

First, at the base of any such wind outflow this momentum equation reduces
to a hydrostatic equilibrium between the inward gravity and 
outward pressure,
\beq
- \frac{1}{P} \, \frac{dP}{dr} \equiv \frac{1}{H_{P}} = \frac{GM}{a^{2} r^{2}} 
\, .
\label{hpdef}
\eeq
Here $a=\sqrt{kT/\mu}$ is the
isothermal sound speed, with $k$ Boltzmann's constant and $\mu$ the
mean molecular weight, 
and we have used the ideal gas law $P = \rho a^{2}$ to obtain an 
expression for the required local pressure scale height $H_{P}$.

The transition to a wind outflow occurs at some radius $R$
where the flow speed becomes supersonic, i.e. $v(R) = a$.
In massive-star winds, for which the temperature is typically close to the
stellar effective temperature, the sound speed $a \approx 20$~km/s,
which is much less than the surface escape speed,
$v_{esc} = \sqrt{2 GM/R} \approx 600-1000$~km/s.
This implies that from the sonic point outward, i.e. from $r>R$, gas
pressure plays almost no role in maintaining the outward
acceleration against gravity, reducing the momentum equation to
\beq
 v \frac{dv }{dr} \approx
 - \frac{GM}{r^{2}} 
 + g_{out}
 ~~ ; ~~ r \ge R
 \, .
\eeq

Integration from this surface radius to infinity then immediately
gives an expression for the required work per unit mass,
\beq
\int_{R}^{\infty} g_{out} \, dr \approx
\frac{v ({\infty})^{2}}{2} + \frac{GM}{R} =
\frac{v_{\infty}^{2}}{2} + \frac{v_{esc}^{2}}{2}
\, .
\eeq
this ignores both the internal and kinetic energy at the sonic point,
since these each are only of order $a^{2}/v_{esc}^{2} \approx 10^{-3}$ 
relative to the terms retained.
For a wind with mass loss rate ${\dot M}$, the global rate of energy
expended is then
\beq
L_{wind} =
{\dot M} \left [ \frac{v_{\infty}^{2}}{2} + \frac{GM}{R}
\right ] \, .
\eeq
The marginal case in which the wind escapes with vanishing terminal 
flow speed, $v_{\infty} = 0$, defines a minimum energy rate for 
lifting material to escape, $L_{min} = {\dot M} GM/R$.
For a given available interior luminosity $L$, this thus implies a maximum 
possible, energy-limited mass loss rate
\beq
{\dot M_{tir}} = \frac{L}{GM/R} = 3.3 \times 10^{-8} \,
\frac{M_{\odot}}{{\rm yr}} \, 
\left [ \frac{L}{M/R} \right ]
\, ,
\label{mdtir}
\eeq
where the latter expression provides a convenient evaluation when the
quantities in square brackets are written in solar units.
The subscript here refers to reduction or ``tiring'' of the radiative
luminosity as a result of the work done to sustain the outflow against
gravity (Owocki \& Gayley 1997).
Even the most extreme massive-star steady winds, e.g. from  WR stars,
are typically no more than a few percent of this energy limit;
but, as discussed further below, the mass loss  during LBV giant eruptions 
can approach this order. 

\subsection{Internal Energy and Virial Temperature}
\label{sec-tintscl}

Although gas pressure is not well-suited to driving a large steady
mass loss from the stellar surface, it is generally the key to supporting
the star against the inward pull of gravity.
The associated pressure scale height is given locally by eqn.
(\ref{hpdef}), which applied at the surface radius $r=R$ gives
\beq
\frac{H_{p}}{R} = \frac{2 a_{eff}^{2}}{v_{esc}^{2}} \approx 10^{-3} \, ,
\eeq
where the latter scaling applies for a surface sound speed set by the
stellar effective temperature, $a_{eff} \approx \sqrt{kT_{eff}/\mu}$.

However, for the stellar interior, the pressure drops from its central
value to nearly zero at the surface, representing an average scale
length $H_{p} \approx R$. This thus implies a characteristic interior 
sound speed $a_{int} \approx v_{esc} $, and a characteristic interior 
temperature
\beq
T_{int} \approx \frac{GM\mu}{kR} 
\approx 1.4 \times 10^{7} \, K \, \frac{M}{R}
\, ,
\eeq
where the ratio $M/R$ is in solar units, with $\mu \approx 10^{-24}$~g, 
roughly appropriate for fully ionized material of solar composition.
This characteristic interior temperature can also be derived from the 
standard ``virial theorum'' result that the stellar internal thermal energy 
is half the gravitational binding energy, implying a negative net
energy that keeps a star gravitationally bound. 

But in the context of mass loss, it means that for gas pressure to have a 
sufficient internal energy to overcome gravity requires a temperature
that is only a factor two
larger than the typical interior value. 
In the solar corona, maintaining temperatures near this escape value 
does allow a pressure-driven solar wind, but this is only possible because 
the low density keeps the wind optically thin,
with thus limited radiative cooling.
For the much higher mass loss rates inferred for hot-star winds,
the much higher density means that radiative cooling would prohibit ever
reaching temperatures much above $T_{eff} \approx T_{int}/1000$.
Thus, as noted above, gas pressure is simply not a viable mechanism 
for driving a dense, steady surface wind.

\subsection{Gas-Pressure-Driven Expansion in Dynamical Explosions}

On the other hand, gas pressure is indeed the primary driving
mechanism for propelling the expansion from supernovae explosions.
In this case, dynamical collapse of the stellar core of mass 
$M_{core} \approx M_{\odot}$ down to a radius characteristic of a neutron
star or black hole, i.e. $R_{ns} \approx 10$~km/s, releases an energy 
\beq
\Delta E \approx \frac{GM_{c}^{2}}{R_{ns}} \approx 10^{53} {\rm erg}
\eeq
which is of order $10^{4}$ higher than the entire binding energy of
the entire stellar envelope,
$E_{g} \approx GM^{2}/R$.
Transfer of just ca. 1\% of this collapse energy 
can thus suddenly heat the surrounding stellar
envelope to a temperature  up to {\em hundred times}
the equilibrium (virial) value, 
with an associated sound speed $a_{sn}$ up to ten times the gravitational 
escape speed.
On a short dynamical time scale, $R/a_{sn}$, of order a few minutes,
the associated large gas pressure then drives an accleration of the full
envelope mass ($\sim 10 M_{\odot}$) to a free expansion at 
speeds $v_{exp} \approx a_{sn}$, typically several thousand km/s.

The radiation generated by such SN explosions escapes on a somewhat longer
time, with light curves typically peaking a few days after the initial
explosion.
But this is still significantly shorter than the characteristic time for LBV giant
eruptions, which apparently can last for several years.
Moreover, the expansion speeds inferred for LBV ejecta are typically a
few hundred km/s, comparable to stellar escape speeds, and much less
than the thousands of km/s typical for the initial expansion of
supernovae.
Thus, rather than a dynamical explosion wherein the gas overpressure
simply overwhelms the binding from stellar gravity, it
seems more likely that LBV eruptions may represent a quasi-controlled
outburst, induced perhaps by an enhanced radiative brightening that leads to
an outward radiative acceleration that exceeds gravity.

\section{Radiatively Driven Mass Loss}

\subsection{Radiative Acceleration and the Eddington Limit}

The force-per-unit mass imparted to material from interaction with
radiation depends on an integration of the opacity and radiative flux 
over photon frequency $\nu$,
\begin{equation}
{\bf g}_{rad} = \int_{0}^{\infty} \, d\nu \, \kappa_{\nu} {\bf
F}_{\nu}/c \, 
\equiv   \kappa_{F} {\bf F}/c
\, ,
\label{gnuint}
\end{equation}
with $c$ the speed of light, and the latter equality defining the
{\em flux-weighted} opacity $\kappa_{F}$ in terms of the bolometric
radiative flux ${\bf F}$.

In general the opacity $\kappa_{\nu}$ includes both broad-band 
continuum processes -- e.g. Thomson 
scattering of electrons, and bound-free or free-free absorption -- 
and bound-bound transitions associated with line absorption and/or 
scattering.
As discussed in \S 3.3, bound-bound opacity is most effective in near-surface 
layers where expansion from a not-too-dense wind can partially desaturate the 
strongest lines.
But in a static envelope and atmosphere, the reduction in flux $F_{\nu}$ 
in such saturated lines keeps the associated line-force small,
and so in most regions of a steller envelope the overall radiative acceleration
is set by continuum processes like electron scattering and bound-free 
or free-free absorption.

In spherical symmetry, both the radial flux $F=L/4\pi r^{2}$ and
gravity $g=GM/r^{2}$ have similar inverse-square dependence on radius $r$, 
which thus cancels in the ratio of
radiative acceleration to gravity.
In terms of the electron scattering opacity,
$\kappa_{e} \approx 0.34~cm^{2}/g$, 
this ratio has the scaling
\begin{equation}
\Gamma \equiv \frac{g_{rad}}{g} 
= \frac{\kappa_{F} L}{4\pi GMc} 
= 2.6 \times 10^{-5} \, {\kappa_{F} \over \kappa_{e}} \, 
{ L \over L_{\odot} } \, {M_{\odot} 
\over M}
\, .
\label{gamdef}
\end{equation}
When the opacity $\kappa$, radiative luminosity $L$, and mass $M$ 
are all fixed, then $\Gamma$ is constant.
But, as discussed below, there are various circumstances in which this is 
not the case.

For pure electron scattering, with $\kappa_{F} = \kappa_{e}$,  eqn. 
(\ref{gamdef}) just gives the classical Eddington parameter $\Gamma_{e}$ 
$=\kappa_{e} L/4\pi GMc$.
Because stellar luminosity generally scales with a high power of the 
stellar mass, i.e. $L \propto M^{3-4}$ (see \S 3.2), massive stars with 
$M > 10 M_{\odot}$\ generally have electron Eddington parameters 
of order $\Gamma_{e} \approx 0.1-1$.
Indeed,  $\Gamma_{e} \equiv 1 $ defines the {\it Eddington limit}, for 
which the entire star would formally become unbound, 
at least in this idealized model  of 1-D, spherically symmetric, 
radiative envelope.

However, because the reversal of gravity formally extends to
arbitrarily deep, dense layers of the stellar envelope, any outward
mass flux that might be initiated would require a very large
mechanical luminosity, and thus would be well above the energy,
photon-tiring limit given in eqn. (\ref{mdtir}).
As such, exceeding the Eddington limit
does not represent an appropriate condition for the  steady-state mass loss 
characteristic of a 
stellar wind, since that  requires an outwardly increasing radiative force 
that goes from being less than gravity in a bound stellar envelope to exceeding 
gravity in the outflowing stellar wind.
The discussion below summarizes how the necessary force 
regulation can still occur through line-desaturation for line driving 
(\S 3.3), and through porosity of spatial structure for continuum driving
(\S 3.5).

But first let us briefly review the key scalings of stellar structure 
that lead massive stars to be so close to this fundamental Eddington limit.

\subsection{Stellar Structure Scaling for Luminosity vs. Mass}

The structure of a stellar envelope is set by the dual
requirements for momentum balance and energy transport.
The former is described through the equation for hydrostatic
equilibrium (cf. eqn. \ref{hpdef}), modified now to account for a 
factor $1-\Gamma$ reduction in the effective gravity,
due to the radiation force.
Following the same approach as in \S\ref{sec-tintscl}, 
this thus now implies a characteristic interior temperature that scales as
\beq
T 
\sim \frac {M 
(1-\Gamma)}{R}
\, .
\label{tmbr}
\eeq

Through most of the stellar envelope, the energy flux 
$F = L/4 \pi r^{2}$ is tranported by
diffusion of radiative energy density $U_{rad} \sim T^{4}$,
\beq
F  = - \frac{1}{\kappa \rho c} \, \frac{dU_{rad}}{dr}
\, ,
\eeq
which implies the dimensional scaling
\beq
L \sim \frac{R^{4} T^{4}}{M} 
\, .
\eeq
When combined with eqn. (\ref{tmbr}) for the interior
temperature, we see that the {\em radius cancels} in the scaling of
luminosity, yielding
\beq
L \sim 
{M^{3} \, 
(1-\Gamma)^{4}}
\, .
\label{mlumscl}
\eeq
Quite remarkably, this scaling does not depend explicitly
on the nature of energy generation in the stellar core, 
but is strictly a property of the envelope structure\footnote{
Of course, this simple one-point scaling relation
does have to be modified to accomodate gradients in the molecular
weight as a star evolves from the zero-age main sequence, 
and it breaks down altogether in the coolest stars (both giants and
dwarfs), for which convection dominates the envelope energy transport.}.

Figure 1 shows a log-log plot of the resulting variation of luminosity
vs. mass.
For low-mass stars, it implies a strong $L \sim M^{3}$ scaling, 
but as this forces stars to approach the Eddington limit, 
the $1-\Gamma$ term acts as a strong repeller away from that limit, 
causing a broad bend toward a linear asymptotic
scaling, $L \sim M$. 

Formally, this scaling suggests it is in principle possible to have
stars with arbitrarily large mass, approaching arbitrarily close to the
Eddington limit.
But surveys of dense young clusters are providing
increasingly strong evidence for a sharp cutoff in the stellar
mass distribution at about $M \approx 150-200 ~ M_{\odot}$
(Oey and Clarke 2005; Kim et al. 2006).

Note that this inferred upper mass limit corresponds closely to
the center of the bend region in fig. 1. 
This is just somewhat beyond the transition, at $\Gamma \approx 1/2$, to where
radiation plays the dominant role in supporting the star against gravity, 
implying a radiation pressure that is greater than gas pressure,
$P_{rad} > P_{gas}$.
Somewhat analagous to having a heavier fluid support a
lighter one, such a configuration may be subject to various kinds
on intrinsic instabilities, leading to spatial clumping and/or
the brightness
variations that trigger LBV eruptions
(Spiegel \& Tau 1999; Shaviv 1998, 2000, 2001).
The large associated LBV mass loss of such near Eddington stars
thus could play a key role in setting the stellar upper mass limit.

\begin{figure}[t]
\begin{center}
\includegraphics[width=5in]{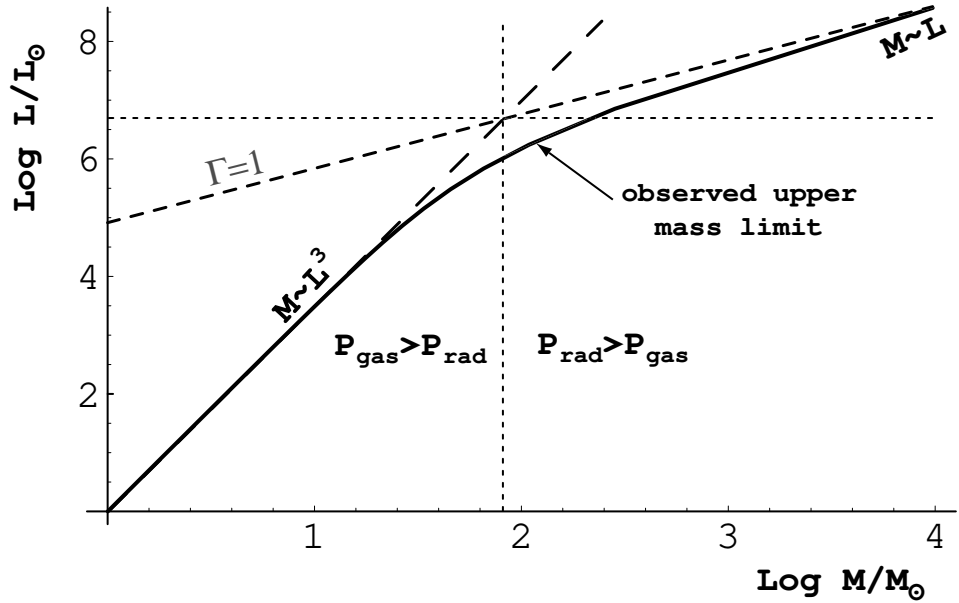} 
\caption{
Log-log plot of the scaling of stellar luminosity $L$ vs.
mass $M$ implied by the simple relation (\ref{mlumscl}).
}
\end{center}
\label{fig1}
\end{figure}

\subsection{Line-Driven Stellar Winds}

%
The resonant nature of line (bound-bound) absorption leads to an opacity that is 
inherently much stronger than from free electrons.
For example, in the somewhat idealized, optically thin limit that all 
the line opacity could be illuminated with a flat, unattenuated 
continuum spectrum with the full stellar luminosity, the total line-force 
would exceed the free-electron force by a huge factor, of order 
$Q \approx 2000$
(Gayley 1995).
For massive stars with typical electron Eddington parameters within a 
factor two of unity, $\Gamma_{e} \approx 1/2$,
this implies a net outward line acceleration that could be as 
high as $\Gamma_{lines} \approx Q \Gamma_{e} \approx 1000$ 
times the acceleration of gravity!

Of course, this does not generally occur in practice because of the 
self-absorption of the lines.
For a single line with frequency-integrated opacity $
\kappa_{q} = q \kappa_{e}$, 
the reduction in the optically thin line-acceleration $q \Gamma_{e}$
can be written as
\beq
\Gamma_{line} \approx q \Gamma_{e} \, 
{ 1 - e^{-qt} \over q t} 
\label{gline}
\, ,
\eeq
where $t \equiv \kappa_{e} \rho c /(dv/dr)$ is the Sobolev 
optical depth of a line with unit strength, $q=1$ 
(Sobolev 1960; Castor, Abbott \& Klein 1975, hereafter CAK).
Within the standard CAK
line-driven wind theory, the number distribution $N$ of spectral lines 
is approximated as a  power law in 
line strength
$ q \, d N/dq = [1/\Gamma(\alpha)] (q/Q)^{\alpha-1}$, 
where the CAK power index $\alpha \approx 0.5-0.7$
(and $\Gamma (\alpha)$ here represents the complete Gamma function).
The associated line-ensemble-integrated  radiation force is then reduced
by a factor 
$1/(Qt)^{\alpha}$ from the optically thin value,
\beq
\Gamma_{lines} = { Q \Gamma_{e} \over (1-\alpha ) (Qt)^{\alpha} }
\propto \left ( { 1 \over \rho} \, { dv \over dr} \right )^{\alpha}
\, .
\eeq

The latter proportionality emphasizes the key scaling of the line-force
with  the velocity gradient $dv/dr$ and {\it inverse} of the density, 
$1/\rho$.
This keeps the line acceleration less than gravity in the dense, 
nearly static atmosphere, but also allows its outward increase above gravity 
to drive the outflowing wind.
The CAK mass loss rate is set by the associated critical density that
allows the outward line acceleration to be just sufficient to overcome the
(electron-scattering-reduced) gravity, 
i.e. with $\Gamma_{lines} \approx 1-\Gamma_{e}$,
\begin{equation}
   {\dot M}_{CAK} = {\alpha \over 1-\alpha } \, { L \over c^{2} } 
   \left [ { Q \Gamma_{e} \over 1-\Gamma_{e} } \right ]^{- 1+ 1/\alpha }
   \, ,
\label{Mdcak}
\end{equation}
where we have used the definition of the mass loss rate 
${\dot M}  \equiv 4 \pi \rho v r^{2}$ and the fact that 
for such a CAK solution,
$v dv/dr \approx g (1-\Gamma_{e})$. 

This last property further yields the characteristic CAK velocity law scaling
$
v(r) \approx v_{\infty} ( 1 - R/r  )^{1/2}
$,
with the wind terminal speed being proportional to the effective surface 
escape speed,
\beq
v_{\infty} \propto v_{eff} \equiv \sqrt{GM(1-\Gamma_{e})/R}
\, .
\eeq

As a star approaches the classical Eddington limit $\Gamma_{e} 
\rightarrow 1 $, these standard CAK scalings formally predict the mass loss rate 
to diverge as ${\dot M} \propto 1/(1-\Gamma_{e})^{(1-\alpha)/\alpha}$, but 
with a vanishing terminal flow speed $v_{\infty} \propto \sqrt{1-\Gamma_{e}}$.
The former might appear to provide an explanation for the large mass 
losses inferred in LBV's, but the latter fails to explain the 
moderately high inferred ejection speeds, e.g. the  500-800 km/s 
kinematic expansion inferred for the Homunculus nebula of $\eta$~Carinae
(Smith 2002, Smith et al. 2003).

But one essential point is that line-driving could never explain the
extremely large mass loss rates needed to explain the Homunculus
nebulae.
To maintain the moderately high terminal speeds, the
$\Gamma_{e}/(1-\Gamma_{e})$ factor would have to be of order unity.
Then for optimal realistic values $\alpha=1/2$ and $Q \approx 2000$
for the line opacity parameters (Gayley 1995), the
maximum mass loss from line driving is given by (Smith \& Owocki
2006),
\beq
{\dot M} \approx 1.4 \times 10^{-4} L_{6} \, M_{\odot}/yr
\, ,
\eeq
where $L_{6} \equiv L/10^{6} L_{\odot}$. 
Even for peak luminosities of a few times $10^{7} L_{\odot}$
during $\eta$~Carinae's eruption, this limit is still several
orders of magnitude below the mass loss needed to form the Homunculus.
Thus, if mass loss during these eruptions occurs via a wind,
it must be a super-Eddington wind driven by continuum radiation
force (e.g., electron scattering opacity) and not lines
(Owocki, Gayley \& Shaviv 2004, hereafter OGS; Belyanin 1999; Quinn \& Paczynski 1985).

\subsection{Convective Instability of a Super-Eddington Stellar Interior}

Before discussing such continuum-driven winds during periods of
super-Eddington luminosity,
it should first be emphasized that locally exceeding the Eddington limit
need {\it not} necessarily lead to initiation of a mass outflow.
As first shown by Joss, Salpeter, and Ostriker (1972),
in the stellar envelope 
allowing the Eddington parameter $\Gamma \rightarrow 1$ 
generally implies through the Schwarzschild criterion that material 
becomes {\it convectively unstable}.
Since convection in such deep layers is highly efficient, the radiative
luminosity is reduced, thereby lowering the associated radiative Eddington
factor away from unity.

This suggests that a radiatively driven outflow should only be initiated
{\em outside} the region where convection is {\em efficient}.
An upper bound to the convective energy flux is set by
\beq
F_{conv} \approx v_{conv} \, l \, dU/dr \ltwig a \, H \, dP/dr \approx a^3 \rho , 
\eeq 
where $v_{conv}$, $l$, and $U$ are the convective velocity, mixing length,
and internal energy density, and $a$, $H$, $P$, and $\rho$ are the sound speed,
pressure scale height, pressure, and mass density.
Setting this maximum convective flux equal to the total stellar energy 
flux $L/4 \pi r^2$ yields an estimate for the maximum mass loss rate 
that can be initiated by radiative driving,
\beq
{\dot M} \le {L \over a^2 } \equiv {\dot M}_{max,conv} = 
{ v_{esc}^{2} \over 2 a^{2} } {\dot M}_{tir} \, ,
\eeq
where the last equality emphasizes that, for the usual case of a sound
speed much smaller than the local escape speed, $a \ll v_{esc}$, such
a mass loss would generally be well in excess of the photon-tiring
limit set by the energy available to lift the material out of the
star's gravitational potential (see eqn.~\ref{mdtir}).  
In other words, if a wind were to originate from where convection becomes
inefficient, the mass loss would be so large that it would  use all the 
available luminosity to accelerate out of the gravitational potential,
implying that any such outflow would necessarily stagnate at some 
finite radius.
One can imagine that the subsequent infall of material would likely 
form a complex spatial pattern, consisting of a mixture of both 
downdrafts and upflows, perhaps even resembling the 3D cells of 
thermally driven convection.

Overall, it seems that a star that exceeds the Eddington limit is 
likely to develop a complex spatial structure, whether due to local 
instability to convection, to global instability of flow stagnation, 
or to intrinsic compressive instabilities arising from the dominance
of radiation pressure.

\subsection{SuperEddington Outflow Moderated by Porous Opacity}

Shaviv (1998; 2000) has applied these notions of a spatially 
structured, radiatively supported atmosphere to suggest an 
innovative paradigm for how quasi-stationary wind outflows could be 
maintained from objects that formally exceed the Eddington limit. 
A key insight regards the fact that, in a laterally inhomogeneous
atmosphere, the radiative transport should selectively avoid regions of 
enhanced density in favor of relatively low-density, ``porous'' channels 
between them.
This stands in contrast to the usual picture of simple 1D, gray-atmosphere 
models, wherein the requirements of radiative equilibrium ensure that the 
radiative flux must be maintained independent of the medium's optical 
thickness.
In 2D or 3D porous media, even a gray opacity will lead to a flux 
avoidance of the most optically thick regions, much as in
frequency-dependent radiative transfer in 1D atmosphere, wherein the 
flux avoids spectral lines or bound-free edges that represent a 
localized spectral regions of non-gray enhancement in opacity.

A simple description of the effect is to consider a medium in which
material has coagulated into discrete blobs of individual optical
thickness $\tau_{b} = \kappa \rho_{b} l$, where $l$ is the blob scale,
and the blob density is enhanced compared to the mean density of the
medium by a volume filling factor $\rho_{b} / \rho = (\L/l)^{3}$, where
$\L$ is the interblob spacing.
The effective overall opacity of this medium can then be approximated as 
\begin{equation}
   \kappa_{eff} \approx \kappa \, { 1 - e^{-\tau_{b}}  \over \tau_{b}} .
\label{keff}
\end{equation}
Note that in the limit of optically thin blobs ($\tau_{b} \ll 1$) this 
reproduces the usual microscopic opacity  ($\kappa_{eff} \approx \kappa$);
but in the optically thick limit ($\tau_{b} \gg 1$), the effective 
opacity is reduced by a factor of $1/\tau_{b}$, thus yielding a medium 
with opacity characterized instead by the blob cross section divided by the 
blob mass ($\kappa_{eff} = \kappa/\tau_{b} = l^{2}/m_{b}$).
The critical mean density at which the blobs become optically thin is 
given by $\rho_{o} = 1/\kappa h$, where $h=\L^{3}/l^{2}$ is 
characteristic ``porosity length'' parameter.
A key upshot of this is that the radiative acceleration in such a 
gray, but spatially porous medium would likewise be reduced by a 
factor that depends on the mean density.

More realistically, it seems likely that structure should occur 
with a range of compression strengths and length scales.
Noting the similarity of the single-scale and single-line correction 
factors (cf. eqns. \ref{gline} and \ref{keff}), let us draw upon an analogy 
with the power-law distribution of line-opacity in the  standard CAK model 
of line-driven winds,
and thereby consider a {\it power-law-porosity} model in which the 
associated structure has a broad range of porosity length $h$. 
As detailed by OGS, 
this leads to an effective Eddington parameter that scales as
\beq
\Gamma_{eff} \approx \Gamma \left ( { \rho_{o} \over \rho } \right 
)^{\alpha_{p}} ~~ ; ~~ \rho > \rho_{o}
\, ,
\eeq
where $\alpha_{p}$ is the porosity power index (analogous to the CAK 
line-distribution power index $\alpha$),
and 
$\rho_{o} \equiv 1/\kappa h_{o}$, with $h_{o}$ now the 
porosity-length associated with the {\em strongest} 
(i.e. most optically thick) clump.

In rough analogy with the ``mixing length'' formalism of stellar 
convection, let us assume this porosity length $h_{o}$ scales with
gravitational scale height $H \equiv a^{2}/g$. 
Then the requirement that $\Gamma_{eff}=1$ at the wind sonic point 
yields a scaling for the mass loss rate scaling with luminosity.
For the canonical case $\alpha_{p}=1/2$, this takes the form
(OGS),
\beqa
{\dot M}_{por}
&\approx& 4 (\Gamma - 1) \, {L \over a c} \, {H \over h_{o}}
\label{Mdporab}
\\
&\approx& 0.004 (\Gamma-1) \, {M_{\odot} \over {\rm yr}} \, {L_{6} \over a_{20} } \, {H \over h}
\, .
\label{Mdporc}
\eeqa
The second equality gives numerical evaluation in terms of characteristic 
values for the sound speed $a_{20} \equiv a/20$~km/s
and luminosity $L_{6} \equiv L/10^{6} L_{\odot}$.
Comparision with the CAK  scalings (\ref{Mdcak}) for a line-driven wind
shows that the mass loss can be substantially higher from a super-Eddington
star with porosity-moderated, continuum driving.
Applying the extreme luminosity 
$L \approx 20 \times 10^{6} \, L_{\odot}$ estimated for the 1840-60
outburst of eta Carinae,
which implies an Eddington parameter $\Gamma \approx 5$, 
the derived mass loss 
rate for a canonical porosity length of $h = H$ is 
${\dot M}_{por} \approx 0.32 M_{\odot}/yr$, 
quite comparable to the inferred average $\sim 0.5 M_{\odot}/yr$ 
during this epoch.
%

Overall, it seems that, together with the ability to drive quite fast 
outflow speeds (of order the surface escape speed), the extended porosity 
formalism provides a promising basis for self-consistent dynamical modeling 
of even the most extreme
mass loss outbursts of Luminous Blue Variables, namely those that, 
like the giant 
eruption of $\eta$~Carinae, approach the photon tiring limit.

\begin{figure}[t]
\begin{center}
\includegraphics[width=5.5in]{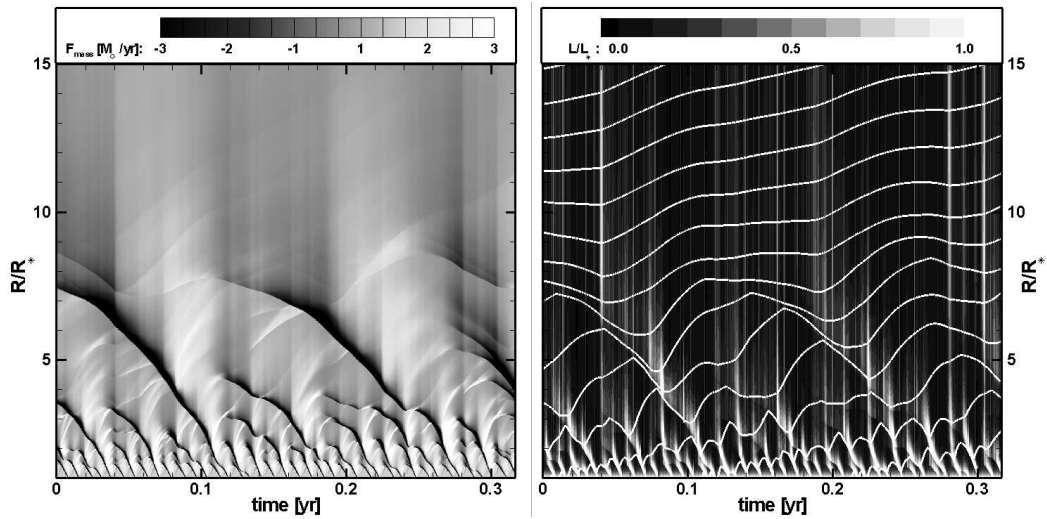} 
\caption{
Grayscale plot of radius and time variation of mass flux (left) and 
luminosity (right) in time-dependent simulation of super-Eddington wind 
with porosity-mediated base mass flux above the photon tiring limit.
The light contours on the right trace the height progression of fixed 
mass shells.
}
\end{center}
\label{fig2}
\end{figure}

\subsection{1D Simulation of Continuum-Driven Winds above the
Photon-Tiring Limit}

For porosity models in which the base mass flux {\em exceeds} 
the photon tiring limit, recent numerical simulations 
(van Marle et al. 2007; see also poster I-49 in these proceedings) 
have explored the nature of the resulting complex pattern of infall 
and outflow.
Despite the likely 3D nature of such flow patterns, 
to keep the computation tractable, this initial exploration assumes 1D 
spherical symmetry, though now allowing a fully time-dependent density 
and flow speed.
The total rate of work done by the radiation on the
outflow (or vice versa in regions of inflow) is accounted for by a
radial change of the radiative luminosity with radius,
\beq
{dL \over dr} 
= - {\dot m} g_{rad}
= - \kappa_{eff} \, \rho v L/c
\, ,
\label{dldr}
\eeq
where ${\dot m} \equiv 4 \pi \rho v r^{2}$ is the local mass-flux at
radius $r$, which is no longer a constant, or even monotonically positive,
in such a time-dependent flow.
The latter equality then follows from the definition (\ref{gnuint}) 
of the radiative acceleration $g_{rad}$ for a gray opacity
$\kappa_{eff}$, set here by porosity-modified electron scattering.
At each time step, eqn. (\ref{dldr}) is integrated from an assumed
lower boundary luminosity $L(R)$ to give the local radiative 
luminosity $L(r)$ at all radii $r > R$.
Using this to compute the local radiative acceleration,
the time-dependent equations for mass and momentum conservation
are evolved forward to obtain the time and radial variation of 
density $\rho (r,t)$  and flow speed $v (r,t)$. 
(For simplicity, the temperature is fixed
at the stellar effective temperature.)
The base Eddington parameter is $\Gamma=10$,
and the analytic porosity mass flux is 2.3 times the tiring limit.

Figure 2 illustrates the flow structure as a function of radius (for
$r=1-15~R$) and time (over an arbitrary interval long after the initial
condition, set to analytic steady porosity model ignoring photon
tiring).
The left panel grayscale shows the local mass flux, in $M_{\odot}/$~yr,
with dark shades representing inflow, and light shades outflow.
In the right panel, the shading represents the local luminosity in
units of the base value, $L(r)/L(R)$, ranging from zero (black) to one 
(white); in addition, the superposed lines represent the radius and
time variation of selected mass shells.

Both panels show the remarkably complex nature of the flow,
with positive mass flux from the base overtaken by a hierarchy of
infall from stagnated flow above. 
However, the re-energization of the radiative luminosity from this
infall makes the region above have an outward impulse.
The shell tracks thus show that, once material reaches a radius 
$r \approx 5 R$, its infall intervals become ever shorter, allowing it
eventually to drift outward.
The overall result is a net, time-averaged mass loss through the
outer that is very close to the photon-tiring limit, with however a
terminal flow speed  $v_{\infty} \approx 50$~km/s 
that is substantially below the 
surface escape speed $v_{esc} \approx 600$~km/s.

These initial 1D simulation thus provide an interesting glimpse into this 
competition below inflow and outflow. 
Of course, the structure in more realistic 2D and 3D models may be even
more complex, and even lead itself to a highly porose medium.
But overall, it seems that one robust property of such super-Eddington
models may well be mass loss that is of the order of the photon tiring
limit.

\section{Conclusion}

The basic conclusion of this review is that the extreme mass loss in
giant eruptions of LBV stars seems best explained by quasi-steady,
porosity-moderated, continuum-driven stellar wind during episodes of 
super-Eddington luminosity.
The cause or trigger of this enhanced luminosity is unknown, but may
be related to the dominance of radiation pressure over gas pressure in
the envelopes of massive stars.
The mass loss rate in such LBV eruptions is far greater than can be
explained by the standard line-driving for hot-star winds in more quiescent
phases.
In the most massive stars, the cumulative mass loss in such eruptions 
may also dominate over the quiescent wind, and might even be a key factor 
in setting the stellar upper mass limit.
Moreover, since driving by continuum scattering by free electrons
does not directly depend on metalicity, mass loss by LBV eruptions may
remain important in low-metalicity environments, including in the early
universe.
A key outstanding issue, however, is to determine the cause or trigger 
of the luminosity brightenings, including, for example, whether this
might itself depend on metalicity.


\par\vskip \baselineskip

\noindent{\it Acknowledgements.}
This research was supported in part by 
NSF grant AST-0507581.
We acknowledge numerous insightful discussions with
N. Shaviv and R. Townsend.

\begin{discussion}

\discuss{Zinnecker}{ I completely agree with you that the term
``radiation pressure'' is ill-conceived, and we should better use a 
term like ``radiative acceleration''. 
I disagree with you, however, on another point: you were writing $L/M$
proportional to $M^{2}$ for very massive stars, when in reality it
should be proportional to $M$ or even $M^{0.6}$ and approaching a
constant near the Eddington limit; see my poster (\# III-32) or
Zinnecker \& Yorke (2007), ARAA 45, 481.
Or did I misunderstand you?
}
\discuss{Owocki}{Yes, I think there was some misunderstanding. The
luminosity scalings you describe agree well with my simple envelope
structure analysis.
See eqn. (\ref{mlumscl}) and fig. 1.
}

\end{discussion}

\end{document}